\documentclass[twocolumn,3p,times,procedia]{elsarticle}

\usepackage{ecrc}
\usepackage{graphicx}
\usepackage{titlesec}
\usepackage{cite}
\usepackage[labelsep=period]{caption}

\usepackage{algorithmicx,algpseudocode}
\usepackage{algorithm}
\usepackage{amsmath,amssymb,amsfonts}
\usepackage{textcomp}
\usepackage{xcolor}
\usepackage{multirow}

\volume{00}

\firstpage{1}

\runauth{}

\jid{procs}

\CopyrightLine{2015}{Published by Elsevier Ltd.}

\usepackage{amssymb}
\usepackage{amsmath}
\usepackage{multicol}

\usepackage[figuresright]{rotating}
\usepackage{bm}
\usepackage{caption}

\usepackage{fancyhdr}

\fancyheadoffset[RO,EL]{0pt}
\usepackage{geometry}

\begin{document}

\begin{frontmatter}




\dochead{}
\title{
\begin{flushleft}
{\bf \Huge A statistical approach for enhancing security in VANETs with efficient rogue node detection using fog computing}
\end{flushleft}
}
 %

\author[]{\bf \Large \leftline {Anirudh Paranjothi$^*$, Mohammed Atiquzzaman}}

\address{\bf  \leftline {School of Computer Science, University of Oklahoma, Norman, Oklahoma, USA}

}

\begin{abstract}

Rogue nodes broadcasting false information in beacon messages may lead to catastrophic consequences in Vehicular Ad Hoc Networks (VANETs). Previous researchers used either cryptography, trust scores, or past vehicle data to detect rogue nodes; however, these methods suffer from high processing delay, overhead, and False-Positive Rate (FPR). We propose herein Greenshield's traffic model-based fog computing scheme called Fog-based Rogue Nodes Detection (F-RouND), which dynamically utilizes the On-Board Units (OBUs) of all vehicles in the region for rogue node detection. We aim to reduce the data processing delays and FPR in detecting rogue nodes at high vehicle densities. The performance of the F-RouND framework was evaluated via simulations. Results show that the F-RouND framework ensures 45\% lower processing delays, 12\% lower overhead, and 36\% lower FPR at the urban scenario compared to the existing rogue node detection schemes even when the number of rogue nodes increases by up to 40\% in the region.

\end{abstract}


\end{frontmatter}



\section{Introduction}

Recent advancements in wireless technology have brought a significant development in Vehicular Ad hoc Networks (VANETs), which are considered as state-of-the-art technology in Intelligent Transportation Systems (ITSs) in terms of enhancing road safety by reducing the number of accidents and optimizing traffic flow. Dedicated Short-Range Communication (DSRC) provides wireless communication capability that permits vehicles to communicate with each other using Vehicle-To-Vehicle (V2V) and Road-Side Units (RSUs) via Vehicle-To-Infrastructure (V2I) techniques [1, 2]. The vehicles are equipped with On-Board Units (OBUs) for transmitting and receiving messages, including beacon messages. VANETs facilitate vehicles to broadcast beacon messages to disseminate the network state or emergency information to reduce road accidents and traffic congestion [3-5]. However, malicious vehicles acting as legitimate vehicles, also known as rogue nodes, may broadcast malicious information, such as false congestion and collision warning for their own benefits [6, 7]. Rogue node detection plays a crucial role in establishing a secure VANET environment because misleading/false information in beacon messages results in changing the normal behavior of vehicles, which may lead to catastrophic consequences, including vehicle collision [8-10]. Therefore, efficient rogue node detection is crucial in containing network damage.

Previous authors used either cryptography, trust scores, or past vehicle data to detect rogue nodes. Al-Otaibi et al.[11] proposed a cryptography-based Intrusion Detection Scheme (IDS) using fog computing to detect rogue nodes. This scheme was known as Fog-IDS. RSUs act as fog nodes transmit public keys to the vehicles in its communication range. The vehicles use private\textendash public key pairs for encrypting messages before being transmitted to the RSUs. Once the messages are received from the vehicles, the RSUs authenticate the key pairs and broadcast the messages to all other vehicles in the region. However, this approach [11] has high processing delay and overhead in detecting rogue nodes when the RSUs are overloaded or not available in the region. Zaidi et al. [12] presented an IDS to detect rogue nodes based on past vehicle data. Each vehicle collects historical information about all other vehicles in the region. Once sufficient details have been collected, the vehicles utilize their OBUs to combine the data and find rogue nodes in the region. The proposed scheme [12] has limitations, such as high delay and overhead. Ahmad et al. [13] proposed a trust-based scheme called Trust Evaluation and Management (TEAM). The TEAM framework comprises three different trust models for detecting rogue nodes: entity, data, and hybrid-oriented trust models. The entity-oriented model performs better compared to the data-oriented and hybrid-oriented trust models in detecting false data exchanged between vehicles due to the presence of high trustworthy vehicles. However, the framework [13] encounters high delay and FPR in detecting rogue nodes when the number of vehicles increases in the region. 

To address the limitations of the existing rogue node detection schemes, we propose herein a Greenshield’s traffic flow based-statistical framework for VANETs, called Fog-based Rogue Nodes Detection (F-RouND). The F-RouND framework is based on fog computing which dynamically utilizes the OBUs of all vehicles in the region for rogue node detection. The proposed framework employs a twofold process in rogue node detection. First, we use the guard node concept to detect rogue nodes. The guard node is the vehicle with more neighboring vehicles in its communication range than all other vehicles and creates a dynamic fog computing layer by utilizing the OBUs of all the vehicles in the region. The fog computing layer is then used to compare the speed and position of all the vehicles to the detect rogue nodes in the region. Second, the guard node performs a hypothesis test to validate whether the rogue nodes are correctly identified. Upon successful validation, the guard node broadcasts the rogue node information to all the vehicles in the region. The F-RouND framework exploits fog computing, which is characterized by low latency and high bandwidth, and performs computations at the network edge [14, 15].  

The \textit{novelty} of the work proposed herein providing low processing delays and FPR at high vehicle densities. In addition, our F-RouND framework does not depend on any roadside infrastructures, such as RSUs, or trust scores or past vehicle data in rogue node detection. The \textit{difference} between the F-RouND framework and the existing schemes [11-13] is that each vehicle uses its OBU or RSU to detect rogue nodes. RSUs are not uniquely deployed in all VANET regions. The absence of RSUs yields high processing delay and FPR. The OBUs of individual vehicles are resource-constrained and cannot be used to process a large number of beacon messages received from all the vehicles in a region because it results in a high delay. VANETs are highly dynamic in nature. The significant delay associated with the rogue node detection may lead to severe network damage. In the F-RouND framework, the OBUs of all the vehicles are combined while creating the dynamic fog layer, which increases the computation power compared to individual OBUs, resulting in low processing delays and FPR. 

Our \textit{objective} is to reduce the latency in detecting rogue nodes, increase the True Positive Rate (TPR), and decrease the FPR at high vehicle densities.  We considered three existing rogue node detection schemes for comparison: Fog-IDS [11], IDS [12], and TEAM [13]. A trade-off always exists when choosing the appropriate rogue node detection schemes for comparison. Choosing from among the existing rogue node detection schemes in VANETs is a challenging task that should consider factors such as the approaches used in rogue node detection, network size, type of data that will be transmitted, mobility models. Based on the aforementioned factors, the schemes presented in [11-13] are the best-known schemes for rogue node detection and have been used by most researchers in the recent times for comparison.  

The performance of the F-RouND framework was evaluated via simulations using Objective Modular Network Testbed in C++ (OMNET++) and Simulation of Urban Mobility (SUMO) simulators for highway and urban scenarios with up to 4000 vehicles and 40\% rogue nodes. Results showed that the F-RouND framework ensures 45\% lower processing delays, 12\% lower overhead, and 36\% lower FPR at the urban scenario, and 44\% lower processing delays, 10\% lower overhead, and 32\% lower FPR at the highway scenario compared to the [11-13] schemes. Overall, our framework performs up to 38\% better than the existing rogue node detection schemes even when the number of rogue nodes increases by up to 40\% in the region. 

The \textit{contributions} of this study are as follows: 
\begin{enumerate}

\item We proposed a framework that uses statistical techniques and Greenshield's traffic model to detect rogue nodes at all vehicle densities in the urban and highway scenarios. 
\item We introduced the guard node in our framework, which uses an OBU-based fog computing technique to compare and analyze beacon messages from all vehicles in the region and to perform an extensive hypothesis test to accept or reject data. 
\item The proposed framework utilizes only vehicle speed values in beacon messages and does not depend on either trust score, cryptography, or past vehicle data in rogue node detection. 
\item We performed an extensive simulation by creating a dynamic fog layer under varying vehicular and network conditions to determine false information broadcasted in both urban and highway scenarios.

The rest of the paper is structured as follows: related work is discussed in Section II; the system model is discussed in Section III. The working principle of the proposed work to detect rogue is illustrated in Section IV; Section V presents the performance evaluation of the F-RouND framework. The simulation results are discussed in Section VI. Finally, the conclusions and future directions are presented in Section VII.

\end{enumerate}


\section{Related Work}

Security is an important issue in the VANET environment. Thus, detecting rogue nodes broadcasting false information through beacon messages plays a crucial role in establishing a secure environment [16, 17]. In VANETs, messages are broadcasted to neighboring vehicles using either broadcasting or multihop techniques. The users make life-saving decisions based on the information received from other vehicles [18]. Therefore, the messages received from the rogue nodes in dynamic vehicular networks may cause havoc by broadcasting false congestion or accident information to the neighboring vehicles. This section presents an overview of three main techniques used in existing approaches to detect rogue nodes in the region: cryptography,  trust-based schemes, and past vehicle data. 

Arshad et al. [19] proposed a beacon-based trust scheme to detect false messages in VANETs. Initially, the trust values of all vehicles are assigned to be 0; then, based on the data correctness, positive or negative trust values are assigned. Positive and negative trust values represent the normal and abnormal behaviors of the neighboring vehicles, respectively. When the calculated trust of any vehicle reaches a predefined threshold limit known as rogue nodes, the information is broadcasted to all the vehicles in the region. However, [19] suffers from high Packet Loss Ratio (PLR) and FPR. Liang et al. [20] presented a feature extraction algorithm for detecting false messages broadcasted by rogue nodes; this algorithm adopts past vehicle data for training and testing the data received from other vehicles in the network. The feature extraction algorithm comprises two modules: a feature extraction module and a classifier module. The feature extraction module extracts the vehicle flow and position from the received vehicle data and sends the information to the classifier module for verification. Upon receiving the values of the vehicle flow and position, the classifier module compares the received values against the past vehicle data to detect the rogue nodes. This approach [20] has a high delay in data extraction and processing at all vehicle densities. 

Sedelmaci et al. and Ahmed et al. [21, 22] proposed trust-based schemes to detect rogue nodes; RSUs were used to compute trust scores and detect rogue nodes based on the calculated trust score. Zhang et al. and Shams et al. [23, 24] illustrated the rogue node detection mechanism based on the Support Vector Machine (SVM) and Dempster\textendash Shafer Theory (DST) to resist false messages. The frameworks [23, 24] comprise a local trust module and a vehicle trust module. The local trust module uses an SVM-based classifier to detect false messages, while the vehicle trust module uses DST to derive the comprehensive trust value for all vehicles. The results of both the local and trust modules are then combined to find the rogue nodes in the region. However, the approaches [21-24] have low TPR and high overhead at high vehicle densities. 

Mundhe et al. [25] illustrated a cryptographic-based message authentication scheme to detect false messages propagated among vehicles in VANETs. The proposed framework [25] employs a ring signature to generate the public and private keys for each vehicle in the network. When the sender broadcasts a beacon message to the neighboring vehicles, a ring signature gets generated and transmitted along with the message. The receiver accepts this message upon successful verification, which helps identify whether the message is modified in the middle. However, a significant delay is associated with the generation of the private and public keys for each vehicle when the RSUs are overloaded or not available in the region. Yang et al. [26] proposed a machine learning algorithm that verifies whether the data received from the vehicles in the region are valid by comparing them with the past vehicle data using classification and decision tree techniques. This approach [26] suffers from low TPR and high FPR when the number of vehicles increases in the region.

Tripathi et al. [27] and Nandy et al. [28] proposed trust-based intrusion detection frameworks for rogue node detection. The models proposed in [27, 28] assigns a trust value to all the vehicles in the region based on their behavior. The vehicles that either drop or alter the message are considered rogue nodes with negative trust values. Vehicles other than the rogue nodes are trustworthy vehicles with positive trust values. The trust values of the vehicles are maintained in the score table, which gets updated whenever messages are broadcasted by the vehicles. Each vehicle has a copy of the score table containing information about all the vehicles in the region. Thus, the messages received from the rogue nodes are ignored to contain network damage. As such, the frameworks proposed in [27, 28] suffer from high PLR and FPR in detecting the rogue nodes. 

Liu et al. [29] proposed a Bayesian interference-based traffic model to identify false messages in VANETs. Based on the Bayesian approach, the model [29] calculates the likelihood of traffic patterns for the future, which is further used to verify whether events, such as road accidents, reported by the vehicle have happened. If the model identifies the reported event as wrong, the corresponding vehicles that broadcasted the messages are considered as rogue nodes. The information regarding rogue nodes are broadcasted to all vehicles in the region. Manimaran et al. [30] presented a framework for Named Data Networking (NDN)-based VANETs. The framework analyzes the messages received from all other vehicles with past vehicle data and predefined rules to detect the rogue nodes. Once the false messages are detected, they are separated from the genuine messages under various test cases, resulting in trustworthy messages being broadcasted to all other vehicles in the region. In [29, 30], the OBUs of each vehicle were used to verify the messages are valid. The OBUs are highly resource-constrained; hence, this approach is not suitable for highly dense regions such as Manhattan and other downtown environments. 

Zhou et al. [31] proposed a distributed collaborative intrusion detection framework that stores and compares past vehicle data to identify the false messages broadcasted by the rogue nodes in VANETs. The scheme proposed in [31] employs a clustering technique to segregate the vehicles based on the reputation state and behavior; then, the normal driving characteristics are compared with the individual clusters to detect malicious behaviors. The vehicles associated with the clusters exposing malicious behaviors termed rogue nodes are then ignored to contain the network damage. This approach [31] suffers from high processing delays in creating clusters and high PLR at high vehicle densities. 

To overcome the limitations of the existing rogue node detection schemes [11-13, 19-31], we propose the F-RouND framework, which does not depend on either trust scores, cryptography, or past vehicle data, but instead adopts the fog computing technique, Greenshield's traffic flow theory, and the statistical model to detect rogue nodes in the region. The hypothesis test used for validating the result of the guard node, which declares whether the vehicle is rogue, increases the efficiency and performance of our framework compared to existing schemes even when the percentage of rogue nodes in the region is 40\%. In addition, the F-RouND framework does not depend on any roadside infrastructures, including RSUs, in the rogue node detection.

\section{System Model}

In this section, we illustrate the mechanism and models, such as the network model, traffic flow model, and attack model, adopted in this study.

\subsection{Network Model}

In VANETs, vehicles communicate with each other through messages. Vehicles are equipped with various devices, such as GPS, Radar, and OBU, to disseminate speed, position, acceleration, braking status, etc. to the neighboring vehicles [32, 33]. V2V and V2I communication is used to transmit messages using either multihop or broadcasting techniques. Based on the received messages from neighbors, every vehicle adjusts their speed, acceleration, etc. to maintain the network state and behavior. We classify the vehicles in the F-RouND framework into two categories: honest, and rogue nodes. Honest nodes are the vehicles broadcasting genuine messages to the neighboring vehicles, while rogue nodes are those injecting false data before broadcasting it to the network. 

We also employ the guard node concept to detect rogue nodes. The guard node is the most trustworthy vehicle located in the center of the network and creates a dynamic fog layer to compare and analyze the vehicle speed in the beacon messages to identify false messages broadcasted by the rogue nodes in the region. The OBUs of individual vehicles are resource-constrained; thus, guard nodes utilize the OBUs of all the vehicles for creating the dynamic fog layer. Our F-RouND framework works efficiently even if there are multiple guard nodes in the region. However, the adoption of the fog computing technique provides a high computation power to the guard node, resulting in a lower delay and a high scalability at all vehicle densities, including highly dense regions, such as downtown environments. Thus, there is no need for multiple guard nodes to detect rogue nodes in any given circumstances. The working principle of the guard node, including the guard node selection, is presented in Section 4.

\subsection{Traffic Flow Model}

We adopted the concept of Greenshield's traffic flow model to model the traffic flows in urban areas and highways. Greenshield model is a fairly accurate and simple model for predicting the traffic flows observed in real-world scenarios. It works under the assumption of density ($\rho$) and speed of vehicles (\textit S) are negatively correlated [34]. The relationship between speed and density is defined as follows:

\begin{equation} \label{eq5}
 \begin{split}
 S \propto \frac{1}{\rho}
 \end{split}
 \end{equation} 
 
 \begin{equation} \label{eq6}
 \begin{split}
 S = C . \frac{1}{\rho}
 \end{split}
 \end{equation} 
where, C is a constant that depends on the communication range of the vehicle. As the speed and density of the vehicles are negatively correlated, the density increases when the vehicle speed decreases in the region and vice versa. The maximum density is the point at which the speed becomes zero, known as (${\rho_{max}}$),\ and the maximum speed is when the density becomes zero, known as ($S_{max}$). 
 
For example, consider the highway scenario depicted in Fig. 1, where vehicles are traveling at high speed. (i.e., 60--70 mph). Under such conditions, the rogue nodes located in the network create a fake road accident scenario by broadcasting low-speed values in the beacon messages to all the vehicles in the region. The red region is where the rogue nodes are located, while the blue region is where the false messages are being broadcasted from the rogue nodes. Upon receiving the beacon messages from the rogue nodes, the vehicles start slowing down considering a road accident ahead. The green region is where the vehicles have not yet started braking because they are not in the communication range of the rogue nodes. The F-RouND framework uses the Greenshield's traffic flow model-based dynamic fog computing technique to detect the rogue nodes (Section 4). However, in the case of an actual road accident scenario, the majority of the vehicles (i.e., honest nodes broadcast low-speed values in the beacon messages) can be easily identified and ignored. 

\begin{figure}[tbp]
\centering
\includegraphics[width=225pt]{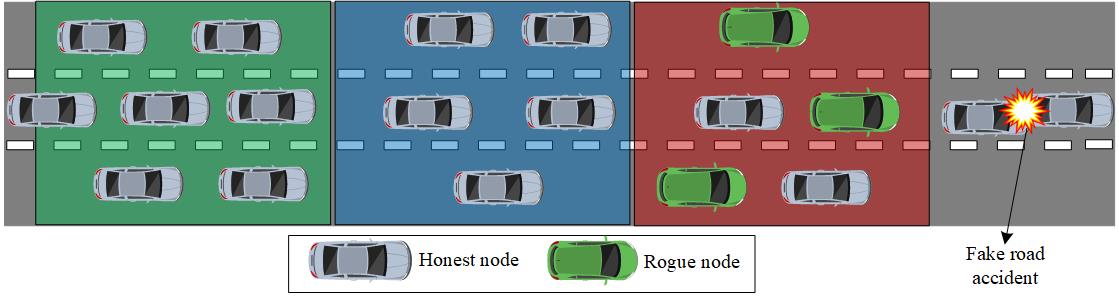}
\caption{An example highway scenario of the traffic flow model.}
\end{figure}

\subsection {Attack Model}

Different types of attacks occur in VANETs. The F-RouND framework addresses the following false information attacks arising from beacon messages:

\textit {False information attack:} Rogue nodes report an arbitrary event by modifying the values in the beacon messages with the vicious intent to cause damage to the network for their benefits. To create a greater impact on the network, the rogue nodes coordinate and collectively modify the speed values in the beacon messages at any time and broadcast low-speed values to the neighboring vehicles in the region to create an illusion of either a traffic congestion or a road accident ahead [35, 36]. Under extreme conditions, the false information messages propagated in VANETs may cause catastrophic consequences, including fatal vehicle collisions. The framework proposed herein effectively monitors the behavior of all the vehicles in the region to the detect rogue nodes, as discussed in Section 4.

\section{Proposed F-RouND Framework}

In this section, we discuss the working principle of the F-RouND framework. The F-RouND framework uses the emerging fog computing technique to detect the rogue nodes providing false information in specific low-speed values by altering the beacon messages in VANETs. The information about rogue nodes are then broadcasted to contain the damage. The proposed framework engages the dynamic fog computing technique to detect the rogue nodes due to its unique characteristics, including low latency and high bandwidth. Moreover, the fog layer is at the proximity of users and performs computations at the network edge [14, 15].

Fog devices share their heterogeneous resources for computation and storage. Moreover, these devices can communicate and cooperate without the intervention of third parties. Fog devices can be either resource-constrained or resource-rich fog nodes with a powerful CPU, large memory, and storage. Individual OBUs are resource-constrained and cannot be used to process a large amount of data. Therefore,  in our F-RouND framework, the OBUs of all the vehicles in the region are considered fog devices and are used for creating a dynamic fog layer at the proximity of the guard node. The dynamic fog layer is located at the network edge. It comprises fog nodes, which include gateways, fog devices, etc.  The fog layer can be static at a fixed location or mobile on moving carriers, such as in the vehicular environment, and is responsible for processing information, such as beacon messages received from the neighboring vehicles, and temporarily storing it or broadcasting it over the network. One of the main advantages of the F-RouND framework is that unlike the existing frameworks [11-13, 19-31], the computation and storage power of the dynamic fog layer increase as the number of vehicles increases in the region, resulting in a lower data processing delay. This is due to the OBUs of all the vehicles being utilized to create the dynamic fog.  

The F-RouND framework employs the guard node concept to detect all rogue nodes in the region. The vehicle with more neighboring vehicles in the communication range is considered as a guard node (Section 4.1). The guard node creates a dynamic fog layer using the OBUs of all the vehicles to compare and analyze beacon messages, specifically the vehicle speeds to detect rogue nodes. In addition to the vehicle speed, beacon messages exchanged between the vehicles also contain information, such as acceleration, braking status, and location. If there is a significant difference in vehicle speeds, the guard node classifies the vehicles as rogue nodes and the hypothesis test is performed to validate whether the rogue nodes are correctly identified. 

If the hypothesis test yields speed values within the acceptance range, then the vehicles are considered as honest nodes; otherwise, the vehicles are highlighted as rogue nodes. Upon successful validation from the hypothesis test, the guard node broadcasts the information of rogue nodes to all the vehicles in the region. The vehicles start ignoring the subsequent beacon messages received from the rogue nodes to contain the damage. One such scenario of our framework is depicted in Fig. 2. The steps involved in the F-RouND framework in the rogue node detection are discussed in the subsections that follow.

\begin{figure}[tbp]
\centering
\includegraphics[width=240pt]{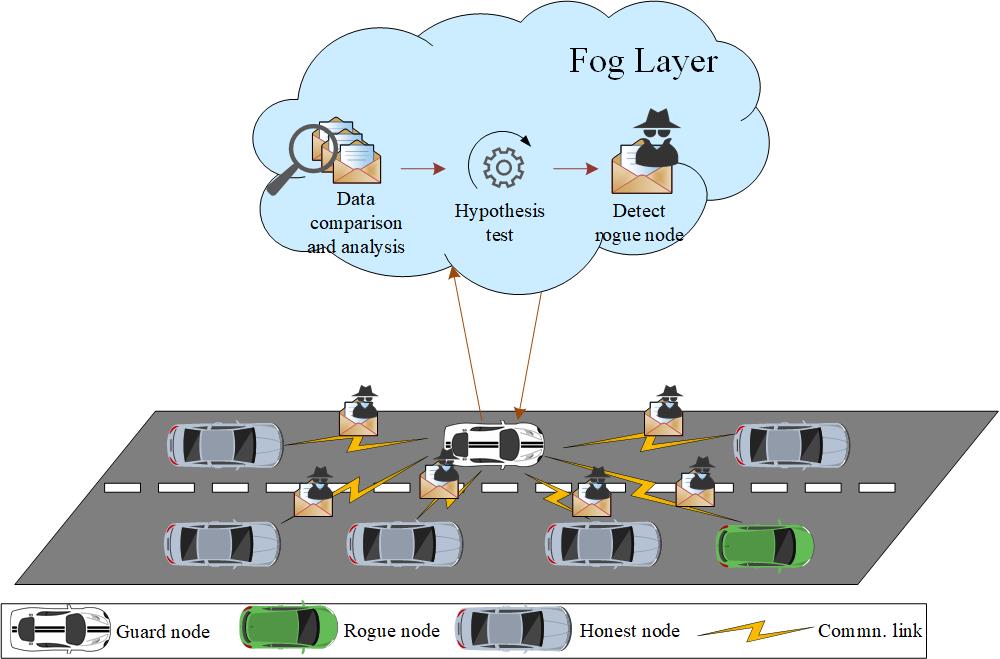}
\caption{Execution scenario of the F-RouND framework in the presence of a rogue node using the dynamic fog computing technique.\label{fig1}}
\end{figure}

\subsection {Guard Node Selection}

Rogue nodes are vehicles broadcasting low-speed values in beacon messages to change the normal behavior of the vehicles for their own benefit. In the F-RouND framework, the guard node analyzes the speed values in the beacon messages received from all vehicles in the region to detect the rogue nodes. The following three assumptions are made in selecting the guard node. First, we assume that the guard node is the most trustworthy vehicle in the network. Thus, the guard node cannot turn out to be a malicious node in any given circumstances. Second, we assume that the center vehicle in the region acts as a guard node because the vehicle in the center has a higher number of neighboring vehicles in its communication range compared to the front and tail-end vehicles. Third, we assume that the total number of vehicles (\textit N) in the region at any given time is at least two. The guard node needs at least two vehicles in the region to compare and analyze the beacon messages to detect the rogue nodes. 

Most often, front-end vehicles are the rogue nodes (Fig. 1) and create an illusion of fake traffic congestion or road accident ahead to trap either one particular vehicle or all the vehicles traveling in the region and create catastrophic consequences, such as vehicle collisions. Moreover, as mentioned in Section 4.1, we assume that the guard node is the central vehicle in the region and cannot be malicious under any circumstances, thus ensuring that center vehicles are not the rogue nodes at any time interval.

Initially, we take the mean of the position vectors of all vehicles (i.e., $P_1$, $P_2$, ...., $P_N$)  to find a unique center point $\zeta$. 

\begin{equation} \label{eq11}
 \begin{split}
 \zeta = \frac{1}{N} \sum_{i=1}^{N} P_i
 \end{split}
 \end{equation}

We calculate the Euclidean distance between $\zeta$ and the position vector of each vehicle and determine the point with the minimum distance from the $\zeta$. Finally, the vehicle located at this point is selected as the guard node,  $G_{veh}$. 

\begin{equation} \label{eq12}
 \begin{split}
 G_{veh} = \arg\min_{P_i \in X} \|\zeta - P_i\|
 \end{split}
 \end{equation} 
where, $X$ = \{$P_1$, $P_2$....,$P_N$\}.
  
 Sometimes, there may be an exceptional case where the point with the minimum distance may not be unique. In such a situation, our F-RouND algorithm will randomly select one vehicle among the minimum distance points as a guard node. 

\subsection {Cooperative Data Collection}

The selected guard node (Section 4.1) collects the data from all the vehicles in the range. The V2V communication technique is used for broadcasting beacon messages. The vehicles share their information using the Greenshield's traffic model discussed in Section 3.2. The vehicles receive beacon messages from all other vehicles in the region; thus, each vehicle knows about all the other vehicles in the region. However, to validate whether the received beacon messages are genuine, the guard node creates a dynamic fog layer by combining the computation resources (i.e., OBUs of all vehicles). In VANETs, the OBUs of individual vehicles are highly resource-constrained and cannot be used to analyze a large volume of data because it will result in a significant delay [37]. Thus, a dynamic fog layer is used for processing the received beacon messages to detect the rogue nodes and to perform an extensive hypothesis test to validate whether the rogue nodes are correctly identified. 

\subsection {Message Format}

In the F-RouND framework, each vehicle broadcasts beacon messages every 100 ms. The format of the beacon message of all the vehicles, except the guard node, is given below:

\,

$B_{msg}$ (\textit{Speed}; $Position$; $Acceleration$; $Density$)

\,

The guard vehicle modifies the existing beacon message format to include information about the rogue nodes in the region. Thus, apart from the usual parameters, the beacon messages from the guard node also include the following:

\,

$B_{msg}$ (\textit{Speed}; $Position$; $Acceleration$; $Density$; $Rlt$; $RID$) \\
where, $ Rlt$ is the result of the hypothesis test and $RID$ is the unique vehicle ID of the rogue nodes discussed in Section 4.5.

\subsection {Speed and Density of Vehicles}

As mentioned in Section 3.2, in the proposed framework (F-RouND), Greenshield's mathematical model is used to model the traffic flow. The guard node calculates the density of vehicles in the region from the received beacon messages from the vehicles in the region and is given as follows:

\begin{equation} \label{eq3}
 \begin{split}
 \rho = B_{msg} \cdot N
 \end{split}
 \end{equation} 
where $B_{msg}$ is the beacon message broadcasted from one vehicle ID and $N$ is the total number of vehicles in the region. The density window size is equal to the transmission and reception ranges of the vehicles in the network. In the F-RouND framework, each vehicle can transmit and receive messages up to 500 m (i.e., the vehicles can communicate with the neighboring vehicles up to 500 m ahead and behind them). Therefore, the communication window of each vehicle, including the guard vehicle, is 1000 m. 
 
The speed and density of the vehicles are negatively correlated (Section 3.2); thus, the relationship between speed and density can be defined as follows:

\begin{equation} \label{eq5}
 \begin{split}
 S = S_{max} -  \frac {\rho}{\rho_{max}} S_{max}
 \end{split}
 \end{equation} 
where $S_{max}$ is the speed of the vehicle when the density is zero and ${\rho_{max}}$ is the maximum density, which is also the point at which the vehicle speed becomes zero.

\begin{table*}[tbp]
\centering
\caption{Types of error and decisions in null hypothesis testing \label{tab1}}
\small
\begin{tabular}{|l|l|l|l|}
\hline
\multicolumn{2}{|l|}{\multirow{2}{*}{}}                                                                                                  & \multicolumn{2}{c|}{\textbf{Null hypothesis ($H_0$)}}                                                                                                    \\ \cline{3-4} 
\multicolumn{2}{|l|}{}                                                                                                                   & \multicolumn{1}{c|}{\textbf{True}}                                      & \multicolumn{1}{c|}{\textbf{False}}                                       \\ \hline
\multicolumn{1}{|c|}{\multirow{2}{*}{\textbf{\begin{tabular}[c]{@{}c@{}}Null hypothesis \textbf{($H_0$)} \\     decision\end{tabular}}}} & \textbf{Accept} & No error                                                                & \begin{tabular}[c]{@{}l@{}}Type II error \\ (False negative)\end{tabular} \\ \cline{2-4} 
\multicolumn{1}{|c|}{}                                                                                                 & \textbf{Reject} & \begin{tabular}[c]{@{}l@{}}Type I error\\ (False positive)\end{tabular} & No error                                                                  \\ \hline
\end{tabular}
\end{table*}

The selected guard node (Section 4.1) creates a dynamic fog layer to compare and analyze speed values in the received beacon messages. If there is a significant speed difference in any of the beacon messages received, the guard node classifies the vehicle that broadcasted the corresponding speed value as a rogue node, and the hypothesis test (Section 4.5) is performed to validate whether the vehicle is rogue.  Once the rogue nodes are identified, the guard node calculates the average density ($\rho_{avg}$) and the average speed ($ S_{avg}$) to perform the hypothesis test.
 
 \begin{equation} \label{eq4}
 \begin{split}
 \rho_{avg} = \frac{1}{N} \sum_{i=1}^{N} \rho_{i}
 \end{split}
 \end{equation} 
 
  \begin{equation} \label{eq7}
 \begin{split}
 S_{avg} = \frac{1}{N} \sum_{i=1}^{N} S_{i}
 \end{split}
 \end{equation} 
 
During the hypothesis test, the guard node compares the average speed  ($ S_{avg}$) with the individual speeds. The vehicles corresponding with the average speed are called honest nodes. In the case where the average speed difference is either high or low, the upper and lower bound values are calculated to decide whether a received speed should be accepted. The validation from the hypothesis test provides better performance of the F-RouND framework, resulting in a higher TPR and a lower FPR compared to the schemes presented in [11-13]. A brief explanation of the hypothesis test is illustrated in Section 4.5.

\begin{figure}[tbp]
\includegraphics[width=221pt, height=11pc]{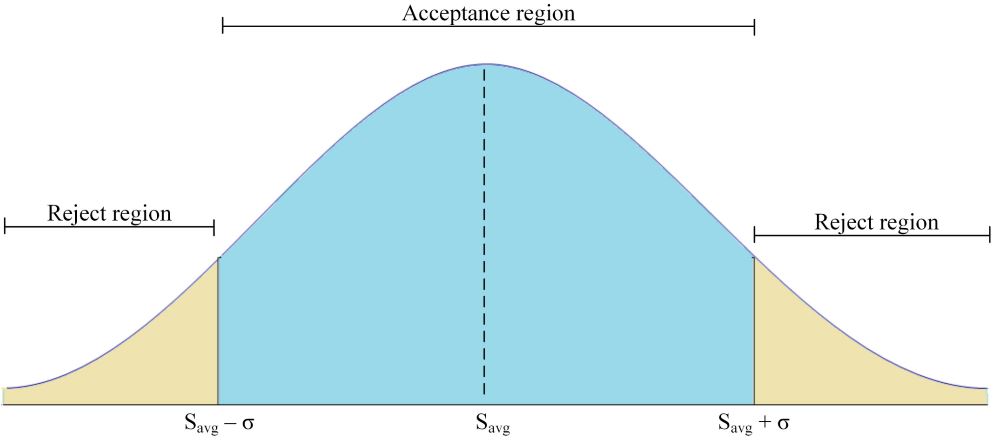}
\caption{Hypothesis test of the F-RouND framework based on the average vehicle speed to determine the acceptance range values.}
\end{figure}

\subsection {Hypothesis Test to Validate the Vehicle Speed}

Hypothesis testing allows a confidence interval to be in a range of values that allows us to accept a claim with a certain confidence. Our framework performs a hypothesis test with the speed values received from all vehicles in the region, allowing the guard node to accept the speeds with a certain confidence. Moreover, hypothesis testing is a commonly used statistical technique when there are two different claims, of which only one claim can be true. In the F-RouND framework, except for the guard vehicle, we have two different claims for all the vehicles in the region (i.e., the vehicle is either honest or rogue). If the vehicle is honest, the guard node accepts the data; otherwise, the vehicle is considered a rogue and the rogue node information is broadcasted over the region. We use the hypothesis test to validate the vehicle speed in the beacon messages, as presented in Fig. 3.

Two hypotheses are involved in the hypothesis testing approach: null hypothesis ($H_0$) and alternate hypothesis ($H_a$). The null hypothesis is the claim that must be tested, while the alternate hypothesis is everything else. If the null hypothesis is accepted, then the alternate hypothesis is rejected and vice versa. In the F-RouND framework, the null hypothesis ($H_0$) is that the speed value received is from an honest vehicle, while the alternate hypothesis  ($H_a$) is that the speed value received is from a rogue node. Two types of error are associated with the hypothesis testing approach: the first type of error (Type I error) occurs when the null hypothesis is wrongly rejected, also known as a false positive; the second type of error (Type II error) occurs when the null hypothesis is wrongly not rejected, also known as a false negative. Table I presents the types of errors and decisions in the null hypothesis testing ($H_0$).

We use standard deviation $(\sigma)$ to calculate the variation in average speed with the received speed values of all vehicles in the region as follows:
 \begin{equation} \label{eq8}
 \begin{split}
 \sigma = \sqrt {\frac{1}{N} \sum_{i=1}^{N} {(S_{avg} - S_{i})}^2}
 \end{split}
 \end{equation} 

The speed values of the vehicles close to the average speed calculated using the guard node result in a low standard deviation, while the speed value of the vehicles that highly from the average speed value results in a high standard deviation. We calculated the confidence interval based on Eq. (9) to determine the acceptance range values. The upper and lower limits of the acceptance region are $S_{avg} - \sigma$ and $S_{avg} + \sigma$. The received speed values of the honest nodes always fall in the acceptance region when the null hypothesis is true (i.e., $S_{avg} - \sigma < S_{avg} < S_{avg} + \sigma$). Therefore, the guard node rejects speed values received outside the acceptance region ($S_{avg} - \sigma > S_{avg} > S_{avg} + \sigma$). The vehicles that broadcast false speed values are considered as rogue nodes. 

Unlike the existing rogue node detection schemes [11-13, 19-31], the F-RouND framework works efficiently for all vehicle densities and all road conditions. For example, in the case of a road accident or traffic congestion, the speed of all the vehicles dropping in the region will bring down the average speed; consequently, the speed values of all honest nodes remain in the acceptance region. The rogue nodes broadcasting false information can be easily detected depending on whether ${S_{avg} > S_{avg} + \sigma}$ or ${S_{avg} < S_{avg} - \sigma}$. Once the rogue nodes are identified, the guard node modifies the existing beacon message format in such a way that the rogue node ID and the result of the hypothesis (either 0 or 1) are embedded in it; then, the guard node broadcasts the information of rogue nodes to all the vehicles in the region. 

 \begin{equation}
  Rlt =
    \begin{cases}
      0; & \text{${S_{avg} - \sigma < S_{avg} < S_{avg} + \sigma}$}\\
      1; & \text{Otherwise}
    \end{cases}       
\end{equation}

\

The beacon message from the guard node includes the following information: \textit ({$Rlt$, $RID$)}. All vehicles in the region start ignoring the beacon messages subsequently received from the rogue nodes to contain the damage.

\subsection {Analysis of the Proposed F-RouND Framework}

In this analysis, we calculated the probability of failure. System failure can occur due to loss of connectivity, insufficient capacity of fog, etc. The probability of system failure $P_{sysfail}$ is calculated as follows:

 \begin{equation} \label{eq13}
 \begin{split}
 P_{sysfail} = \sum_{i=0}^{N, t_{max}}{\left(N, t_{max} \atop i\right)}d_f^i(1-d_f)^{{N,t_{max}}^{-i}}
 \end{split}
 \end{equation} 
 where $N$ is the number of vehicles, $t_{max}$ is the maximum time taken by the vehicles to get connected, and $d_f$ is the probability of success in the fog.  A minimum number of failures leads to the maximum performance of the F-RouND framework.
 
 \subsection {F-RouND Rogue Node Detection Algorithm}

\begin{algorithm}
\caption{: Rogue nodes detection algorithm }\label{alg1}
\textbf{Input:} $G_{veh}$ receives $B_{msg}$ from all vehicles in the region
\newline
\textbf{Output:} $G_{veh}$ broadcasts the rogue node information to all the vehicles in the region

\begin{algorithmic} [1]
\If {($N$ $\geq$ 2)} 
\State Calculate  $\zeta$
\State Calculate Euclidean distance 
\State Assign  $G_{veh}$ 
\Else \, Goto step 21
\EndIf
\State $G_{veh}$ creates a dynamic fog layer from OBUs of all vehicles in the region
\State $G_{veh}$ receives $B_{msg}$ from all vehicles in the region
\For {each  $B_{msg}$  received} 
\State Calculate $\rho_{avg}$
\State Calculate $S_{avg}$
\State Perform hypothesis test to validate each vehicle speed 
\If {$S$ in the acceptance range}
\State Declare the vehicle as honest node
\Else
\State Declare the vehicle as rogue node
\State Store the rogue node id
\EndIf
\EndFor
\State $G_{veh}$ broadcasts rogue node information through $B_{msg}$ 
\State Terminate rogue node detection algorithm

\end{algorithmic}
\end{algorithm}

\begin{figure*}[tbp]
\centering
\includegraphics[width=400pt]{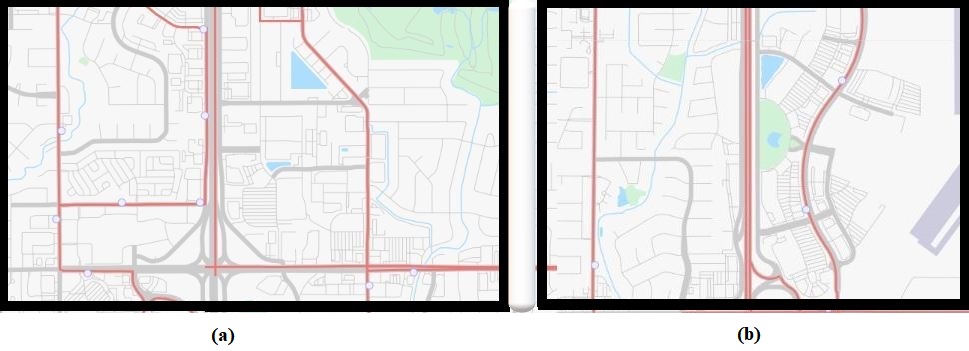}
\caption{Simulated maps of the F-RouND framework from the City of Norman, USA: (a) urban and (b) highway scenarios. \label{fig1}}
\end{figure*}

\section{Performance Evaluation}

This section evaluates the performance of our proposed framework discussed in Section 4. Each analysis is explained in the following subsections.

\subsection {Simulation Setup}

The main objective of our simulation was to evaluate the performance of the F-RouND framework in the presence of rogue nodes in both urban and highway scenarios (Section 4). We used OMNET++ and SUMO simulator to perform the simulations. SUMO is an open-source traffic simulator that provides a trace of vehicle movements, such as vehicle speed, position, and acceleration. at the end of every simulation [38]. SUMO supports OpenStreetMap (OSM) to import real-world road networks, including buildings, water bodies, and traffic lights, for a realistic simulation. The output of the SUMO simulation was given as input to the OMNET++ simulator. OMNET++ is a discrete event simulator that provides a packet loss model, a node deployment model, a node mobility model, and a wireless signal propagation model to measure the network performance [39]. The node deployment and mobility models were used for determining the dynamic placement and movement of vehicles, respectively, while the wireless signal propagation and packet loss models were used for transmitting radio waves and measuring the number of packets dropped in the transmission, respectively. 

To measure the performance of our F-RouND framework, we imported two maps from the city of Norman, United States of America. The first map represents the urban scenario, while the second shows the highway scenario of Norman, as represented in Fig. 4. The vehicles in the urban scenario had a lower mobility compared to those in the highway scenario as the vehicle speed in the urban scenario was limited to 30\textendash 45 mph whereas that in the highway scenario was up to 70 mph. The majority of the vehicles in the network were honest and broadcasted genuine messages to the other vehicles in the network. Thus, to assess the scalability and behavior of the F-RouND framework, we increased the presence of rogue nodes in the network up to 40\%. Table II summarizes the most commonly used parameters used in the simulation.

\begin{table}
\caption{Parameters used in simulartion of the F-RouND framework \label{tab2}}
\centering
\small
\begin{tabular} [htbp] {|c|c|}

\hline

\textbf{Parameters} & \textbf{Values} \\    \hline \hline
Road length & 6 km \\    \hline
Number of vehicles & 500--4000 \\    \hline
Number of lanes & 2 \\    \hline
Vehicle speed &  30--70\\    \hline
Beacon message size & 300 bytes \\    \hline
Transmission range & 500 m\\    \hline
Simulation scenario & Urban and highway\\ \hline
Technique used & Fog computing \\    \hline
Protocol & IEEE802.11p \\    \hline
Simulator used & Omnet++, SUMO \\    \hline

\end{tabular}
\end{table}

All the vehicles in our simulations used the IEEE 1609 Wireless Access in Vehicular Environment (WAVE) protocol/DSRC stack, which builds on IEEE 802.1p WLAN standard and operates on seven reserved channels in the 5.9-GHz frequency band for vehicular communication. Among the seven channels, the DSRC stack has one Control Channel (CCH) responsible for broadcasting critical information, such as beacon messages, and Six Service Channels (SCHs) for broadcasting non critical information. The vehicles continuously adopted CCH to broadcast 300-byte beacon messages to the neighboring vehicles every 100 ms at an application rate of 3 Mbps and SCHs to randomly send 256 byte IP packets at an application rate of 6 Mbps. Our measurements were based on averaging the results obtained from 10 simulations. We varied the number of rogue nodes to be between 5\% and 40\% of the overall vehicles in the network. The total simulation time was 700 s. The rogue node detection process started immediately once all the vehicles entered the road. To provide stochasticity in the simulation, we used randomness in SUMO, which allows the vehicles to enter and exit from a random lane with a variable speed and destination. This eliminates a controlled environment with a predictable outcome, leading to a reality in the simulation.

OMNET++ uses a TCP-based client-server architecture, where OMNET++ acts as a client and SUMO acts as a server. It helped simulate the real streets of the city of Norman by considering the lanes, traffic lights, turns, and other traffic entities. In our F-RouND framework, the threshold value was more dynamic and was calculated for each simulation as it depends on the speed of the vehicles in the region. The low-speed threshold may increase the FPR. However, when the speed threshold was increased, the detection probability decreased because some rogue nodes may have been missed due to the high-speed threshold. Consequently, to detect all the rogue nodes in the region and to reduce the FPR, the speed threshold has to be chosen dynamically based on the simulation scenario. Once the rogue node was detected in the simulation, the guard node changed the corresponding vehicle state in SUMO to a rogue node by sending a message to the OMNET++ client interface, which generated a set of commands and sent them to SUMO for execution, followed by broadcasting the rogue node information to all the vehicles in the region.

\subsection {Performance Metrics}

The simulations were performed based on the equations formulated in Section 4. The number of rogue nodes was increased from 5\% to 40\% to identify how successfully our proposed framework classifies trustworthy vehicles as honest nodes and malicious vehicles as rogue nodes. We considered data processing time, PLR, average throughput, overhead, TPR, and FPR to evaluate the performance of the  F-RouND framework and to compare our results with those of the Fog-IDS, IDS, and TEAM schemes:

\begin{itemize}

\item Data processing time: The time needed by the guard node to compare and analyze the beacon messages to detect rogue nodes in the region.

\item PLR: The ratio of the number of lost packets to the total number of packets sent across a communication channel.

\item Average throughput: Average rate of successfully broadcasted messages across a communication channel. 

\item Overhead: The overhead is the additional information exchanged between the vehicles to detect rogue nodes in the region.

\item True positive rate: The percentage of rogue nodes is accurately detected and classified as rogue nodes.

 \begin{equation} \label{eq13}
 \begin{split}
\text {TPR} = \frac{\text {Number of rogue nodes detected correctly}}{\text{Total number of rogue nodes}}
 \end{split}
 \end{equation}

\item False positive rate: The percentage of honest nodes is incorrectly detected and classified as rogue nodes. 

 \begin{equation} \label{eq13}
 \begin{split}
\text {FPR} = \frac{\text{Number of honest nodes detected incorrectly}}{\text{Total number of honest nodes}}
 \end{split}
 \end{equation}

\end{itemize}

  \begin{figure*}[htbp]
\centering
\includegraphics[width=500pt]{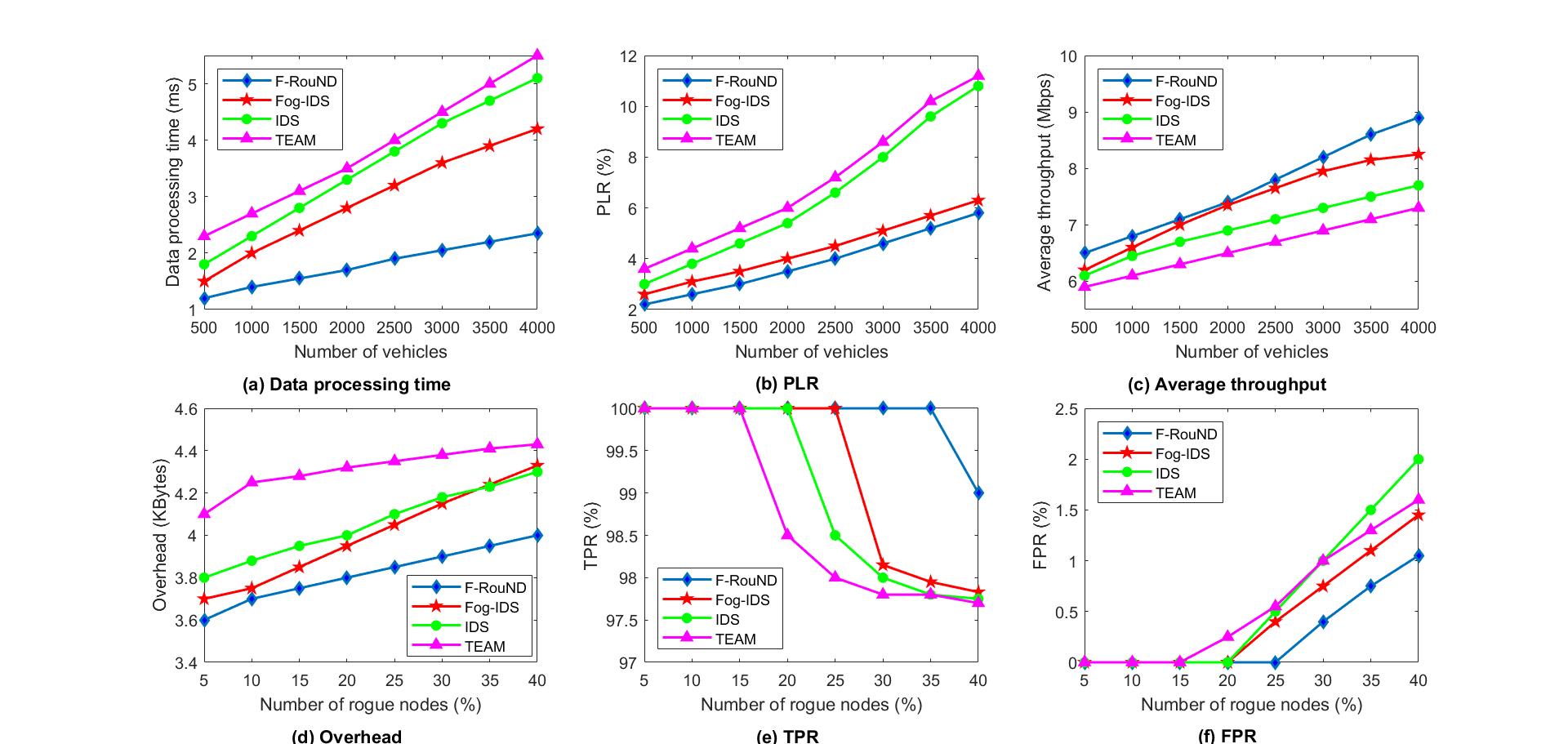}
\caption{Comparison of urban scenarios of the F-RouND framework with Fog-IDS, IDS, and ELIDV schemes: (a) data processing time; (b) PLR; (c) average throughput; (d) overhead; (e) TPR; and (f) FPR. \label{fig3}}
\end{figure*}

\section {Results}

As mentioned in Section 5, we performed a simulation in two parts: the urban and highway scenarios. The vehicles in the urban scenario had a low mobility, while those in the highway scenario had a high mobility. During the simulation, the F-RouND  framework was analyzed based on the equations formulated in Section 4 and the performance metrics illustrated in Section 5. The results are presented in the subsequent sections.

\subsection {Urban Scenario}

\textit{1) Data processing time:} This is the time taken by the guard node to process and analyze the received beacon messages from the neighboring vehicles. When the number of vehicles increased from 500 to 4000, the data processing time increased as the guard node needed to process a large number of beacon messages received from all the vehicles in the region. However, as mentioned in Section 4, the computation power of the guard node increased when the number of vehicles increased in the region because the OBUs of all vehicles utilized in creating the dynamic fog layer resulted in a 45\% lower processing delay at all vehicles densities compared to the schemes presented in [11-13]. In the 4000-vehicle simulation, the data processing time was 43\%, 52\%, and 57\% lower than that in the Fog-IDS, IDS, and TEAM schemes, respectively as shown in Fig. 5a. The results show that our F-RouND framework is efficient, scalable, and can handle high vehicle densities.  

\textit{2) PLR:} The PLR increased when the number of vehicles increased in the region because an increase in the number of vehicles receives a large number of beacon messages from the neighboring vehicle. Thus, the load on the dynamic fog layer increased, which consequently resulted in a packet drop when the load reached the maximum capacity of the dynamic fog. However, the high computation power of the fog resulted in an optimum network capacity even at high vehicle densities in an urban scenario. The PLR was calculated for a number of vehicles ranging from 500 to 4000, as shown in Fig. 5b. In the 4000-vehicle simulation, the PLR was 8\%, 45\%, and 47\% lower than that in the Fog-IDS, IDS, and TEAM schemes, respectively.

\textit{3) Average throughput:} In our F-RouND framework, the average throughput increased when the number of vehicles increased from 500 to 4000 because a large number of beacon messages were successfully broadcasted to all vehicles in the region. This was due to the high scalability of our dynamic fog, low data processing delays (Fig. 5a), and low PLR (Fig. 5b) at high vehicle densities. In the 4000-vehicle simulation, the average throughput in an urban scenario was 9\%, 17\%, and 23\% higher than that in the Fog-IDS, IDS, and TEAM schemes, respectively. The evaluation of average throughput shows the robustness and efficiency of the F-RouND framework, as shown in Fig. 5c. 

\textit{4) Overhead:} The overhead of the F-RouND framework was calculated against the number of rogue nodes, as shown in Fig. 5d, and increased at all vehicle densities due to an extensive hypothesis test needed to validate whether or not the rogue nodes were correctly identified (Section 4). However, unlike existing approaches [11-13], our F-RouND framework does not require any additional information, such as past vehicle data, trust score, and digital signature exchanged between the guard node to detect rogue nodes in the region resulting in 12\% lower overhead in an urban scenario even when the number of rogue nodes increased up to 40\% as shown in Fig. 5d. For example, when the number of rogue nodes was 30\% in the region, the overhead was 10\%, 9\%, and 13\% lower than that of the Fog-IDS, IDS, and TEAM schemes, respectively.

\textit{5) TPR:} TPR was calculated against the number of rogue nodes, as shown in Fig. 5e. When the number of rogue nodes was more than 35\%, the TPR marginally decreased to 99\%. It was difficult to detect the rogue node when the speed variation was gradual. However, to generate either a false congestion scenario or catastrophic consequences, the target rogue node suddenly decreased the speed values. Thus, the F-RouND framework can detect the rogue nodes even at high vehicle densities, resulting in a higher TPR compared to [11-13].  

\textit{6) FPR:} The increase in the FPR increased the rogue nodes in the region and deteriorated the performance of the rogue node detection schemes. One of the main advantages of the F-RouND framework is that the rogue node detection relies only on the speed values in the beacon messages broadcasted by all vehicles in the region without considering any trust scores or past vehicles, resulting in 36\% lower FPR even at 40\% rogue nodes in the region compared to [11-13] schemes. For example, when the number of rogue nodes was 40\% in the region, the FPR was 31\%, 51\%, and 38\% lower than that of the Fog-IDS, IDS, and TEAM schemes, respectively as shown in Fig. 5f.

 \begin{figure*}[tbp]
\centering
\includegraphics[width=500pt]{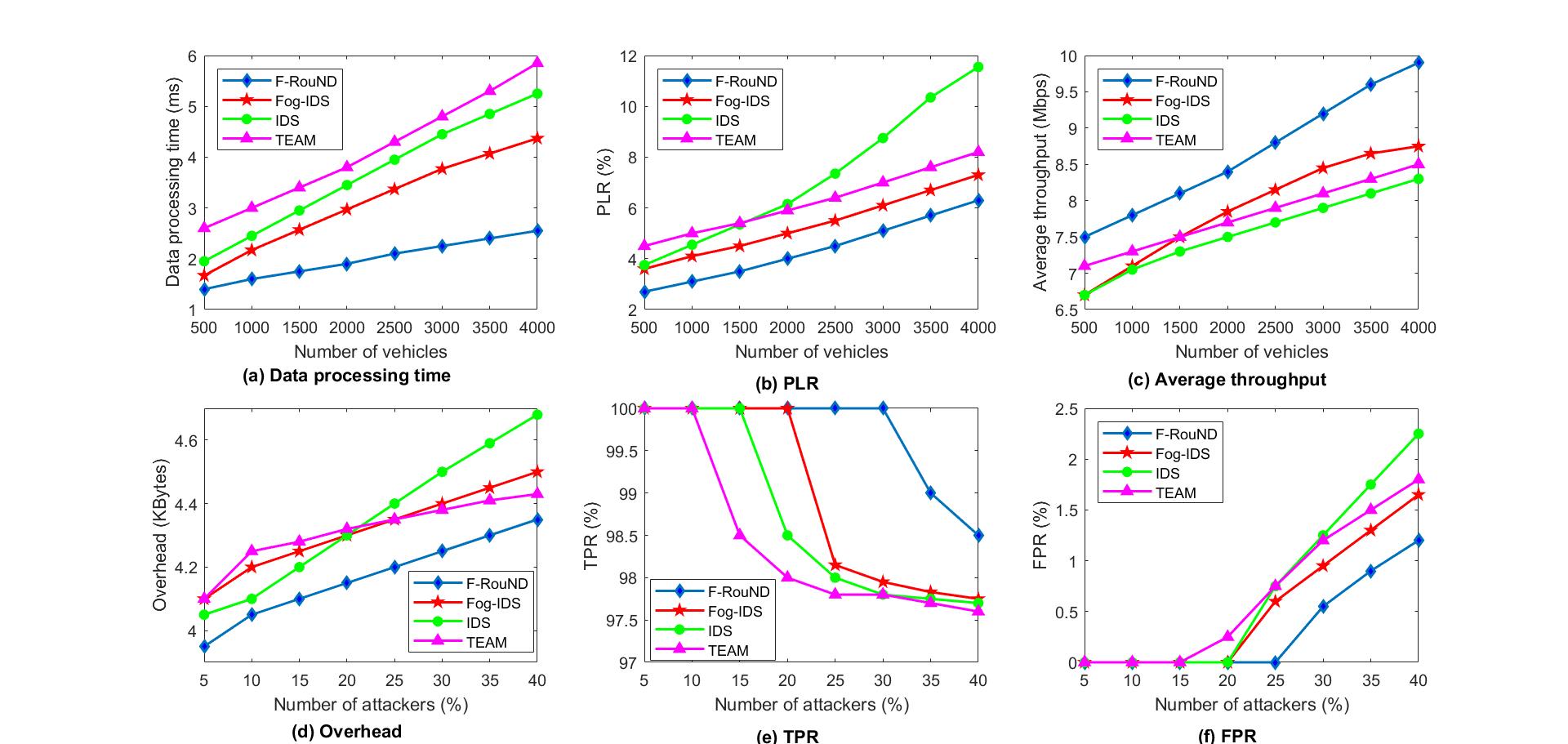}
\caption{Comparison of highway scenarios of the F-RouND framework with Fog-IDS, IDS, and ELIDV schemes: (a) data processing time; (b) PLR; (c) average throughput; (d) overhead; (e) TPR; and (f) FPR. \label{fig3}}
\end{figure*}

\subsection {Highway Scenario}

\textit{1) Data processing time:} Fig. 6a shows the data processing delay for the number of vehicles ranging from 500 to 4000 in a highway scenario. The F-RouND framework showed a relative insensitivity to vehicle counts because the guard node adopted a dynamic fog layer for processing the beacon messages received from neighboring vehicles (Section 4). Therefore, irrespective of the simulation scenario (i.e., either urban or highway scenario), the delay remained similar. The data processing delay at the highway scenario was 44\% lower compared to those in [11-13], resulting in a highly robust and efficient framework. For example, in the 3000-vehicle simulation, the data processing delay was 40\% lower than the Fog-IDS, IDS, and TEAM schemes. 

\textit{2) PLR:}  The PLR of our F-RouND framework in a highway scenario was marginally higher compared to the urban scenario was due to the high mobility of the vehicles, which resulted in a collision of some packets. However, the  PLR of our highway scenario was lower compared to that of the schemes presented [11-13] at all vehicle densities. The PLR was calculated against the number of vehicles and increased for all schemes with increasing number of vehicles. In the 4000-vehicle simulation, PLR was 12\%, 41\%, and 23\% lower than that in the Fog-IDS, IDS, and TEAM schemes respectively, as shown in Fig. 6b.

\textit{3) Average throughput:} Due to a large number of messages being successfully broadcasted to the neighboring vehicles in the dynamic fog region, the average throughput of the F-RouND framework was higher compared to that in the schemes presented in [11-13] in a highway scenario. The average throughput was calculated against the number of vehicles and increased when the number of vehicles increased from 500 to 4000 as shown in Fig. 6c. In the 4000-vehicle simulation, the average throughput was 13\%, 19\%, and 16\% higher than that in the Fog-IDS, IDS, and TEAM schemes, respectively. 

\textit{4) Overhead:} Fig. 6d shows the effect of change in the number of rogue nodes on the overhead in a highway scenario. As mentioned in Section 6.1, the validation of the hypothesis test increased the overhead of our framework when the number of rogue nodes increased in the region. However, the F-RouND framework only relies on speed values in the received beacon messages when detecting rogue nodes. Thus, the overhead of the F-RouND framework in the highway scenario was 10\% lower compared to that in the schemes presented in [11-13]. For a network with 40\% rogue nodes, the overhead was 13\%, 7\%, and 6\% lower than that in the Fog-IDS, IDS, and TEAM schemes, respectively. 

\textit{5) TPR:} The TPR of the F-RouND framework in the highway scenario identified the rogue nodes correctly up to 30\% rogue nodes in the region and slightly decreased to 98.5\% when the number of rogue nodes increased to 40\% as shown in Fig. 6e. It was difficult to detect the rogue nodes broadcasting false information when the speed varied gradually in the received beacon messages. However, to generate either a road accident scenario or catastrophic consequences, the target rogue node suddenly decreased the speed values resulting in a higher TPR compared to that in the Fog-IDS, IDS, and TEAM schemes at all vehicle densities. 

\textit{6) FPR:} An increase in FPR is a critical issue for any rogue node detection schemes as it increases the number of rogue nodes that may lead to severe network damage. The twofold process of the F-RouND framework (i.e., comparison of the speed values in beacon messages) for detecting rogue nodes and validating an extensive hypothesis test to determine whether the rogue nodes are correctly identified or not resulted in 32\% lower FPR compared to the [11-13] schemes at all vehicle densities as shown in Fig. 6f. For a network with 40\% rogue nodes, the FPR was 25\%, 46\%, and 32\% lower than Fog-IDS, IDS, and TEAM schemes, respectively.

\section{Conclusions and Future Directions}
We studied herein the challenges in the existing rogue node detection schemes [11-13, 19-31], including high delay, poor resource utilization, high FPR, and low TPR at high vehicle densities when the number of rogue nodes increases in a region. To address these limitations and provide an efficient scheme to detect the false messages broadcasted over the network, we proposed Greenshield's traffic flow-based fog computing technique called F-RouND for rogue node detection. The F-RouND framework demonstrated the effectiveness of the fog computing technique in determining rogue nodes, even when the number of rogue nodes increases by up to 40\% in the region. 

The simulations were performed based on metrics such as the data processing time, PLR, average throughput, network overhead, TPR, and FPR to evaluate the performance of the proposed framework using OMNET++ and SUMO simulator.  Results showed that the F-RouND framework is scalable, efficient, robust, and performs up to 38\% better than the schemes presented in [11-13]. Moreover, the performance of our extensive hypothesis test in validating the rogue nodes ensured 45\% lower processing delay, 36\% lower FPR and 12\% lower overhead at the urban scenario and 44\% lower processing delay,  32\% lower FPR, 10\% lower overhead at the highway scenario compared to that in the Fog-IDS, IDS, and TEAM techniques [11-13]. The F-RouND framework does not depend on any roadside infrastructures such as RSUs or trust scores or past vehicle data in rogue node detection, which is a major advantage compared to existing rogue node detection schemes.

In the future, we plan to extend this work to detect various attacks, such as Sybil, GPS spoofing, and impersonation attacks. This can be done by simulating the environment of the security attacks and then detecting the malicious information
broadcasted using rogue node detection techniques.

~~~\\
~~~\\







\end{document}